\documentclass{aastex}     %% The manuscript based on AASTeX v5.x
\usepackage{url}
\usepackage{comment}
\usepackage{epstopdf}
\usepackage{gensymb}

\usepackage[maxfloats=40] {morefloats} % add to eliminate "too many unprocessed floats error"

\begin{document}
\title{An Investigation into Exoplanet Transits and Uncertainties}
\shorttitle{Exoplanet Transits and Uncertainties}
\shortauthors{<Y. Ji et al.>}

%% Author and Affilations
\author{Y.\ Ji\altaffilmark{1}}
\and
\author{T.\ Banks\altaffilmark{2}}
\and
\author{E.\ Budding\altaffilmark{3,4,5,6}}
%\affil{UoC, Christchurch, New Zealand.}
\and
\author{M.~D.\ Rhodes\altaffilmark{7}}

\altaffiltext{1}{Munich Re, 20 Collyer Quay {\#}13-01, Singapore 049319}
\altaffiltext{2} {Nielsen, Data Science, 47 Scotts Rd {\#}13-00, Singapore 228233}
\altaffiltext{3}{University of Canakkale, TR 17020, Turkey;}
\altaffiltext{4}{Dept.\ Physics \& Astronomy, UoC, New Zealand.}
\altaffiltext{5}{SCPS, Victoria University of Wellington and}
\altaffiltext{6}{Carter Observatory}
\altaffiltext{7}{BYU, Provo, Utah}

\vspace{2mm} 

%-------------------------------------------
\begin{abstract}

A simple transit model is described along with tests of this model
against published results for 4 exoplanet systems (Kepler-1, 2, 8, and
77).  Data from the {\em Kepler} mission are used.  The Markov Chain
Monte Carlo (MCMC) method is applied to obtain realistic error
estimates.  Optimisation of limb darkening coefficients is subject to
data quality.  It is more likely for MCMC to derive an empirical limb
darkening coefficient for light curves with S/N (signal to noise) above
15.  Finally, the model is applied to {\em Kepler} data for 4 {\em Kepler}
candidate systems (KOI 760.01, 767.01, 802.01, and 824.01) with
previously unpublished results.  Error estimates for these systems are
obtained via the MCMC method.
    
\end{abstract}
%-------------------------------------------


%-------------------------------------------

\section{Introduction}

The last two decades have seen an explosion in the number of planets
confirmed to orbit stars other than the Sun (see Pollacco {\em et al.}
(2006) and Rice (2014) for reviews). To date, the transit method has
been the leading technique for exoplanet detection, with the {\em
Kepler} mission being the major contributor to the transit detections.
The scientific aims of the {\em Kepler} mission were presented by
Borucki {\em et al.}~(2003), while Borucki {\em et al.} (2011) gave an
early summary of results. This has been updated by Rowe {\em et al.}
(2014), Mulally {\em et al.} (2015) and others.
{ Although {\em Kepler} revolutionised the science of
exoplanet transits, it is worth noting that other space missions have
also strongly contributed to the field, for instance, the first
transiting exoplanet detection using a space-based observatory was by CoRot
(Moutou~{\em et al.}, 2013).  We note with anticipation the upcoming
Transiting Exoplanet Survey Satellite (TESS) mission (Ricker~{\em et
al.}, 2010).}

The Kepler Science Center has managed the organization of data for
scientific users, such data being readily available from the NASA
Exoplanet Archive
({http://exoplanetarchive.ipac.caltech.edu/} being the
``NEA Website" or NEA; { see also Akeson~{\em et al.},
2013}). Data collected by {\em Kepler} and other exoplanet research
projects are generally available to researchers and the public.  This
paper made use of data from one such online resource: the NEA, which
also provides tools to manipulate these data as well as summary tables
from previous analyses. We have been interested to investigate the
uncertainties of parameter estimation for fits to the {\em Kepler}
photometric data, in particular the transit regions of light curves.  We
believe it should be standard practice to state clearly uncertainties of
light curve models or fits and how uncertainties are estimated, giving
indications of the reliability of the estimates.  This gains importance
as research moves from the study of individual systems to population
analysis.

The approach followed in this study is as follows: 

\begin{itemize}

	\item{We decided to start with first principles and build up a
model, so that we clearly understood the fitting function being
applied. This was a simple model for single-planet systems in
circular orbits, which was then fitted to folded and binned
individual light curves for 4 systems with published parameter sets. 
We applied five different optimisation techniques, to see if they
settle in the same vicinity of the solution space.  If such fits,
{ or validation tests,} fall
close to the published results for a system, then this should lend
some confidence that our model was implemented correctly.  However,
optimisation techniques generally provide point estimates for fitted
parameters, without clear indications of the quality of those
estimates.}

	\item{To explore the scatter in the fits, we next fitted the
model to multiple transits per system, and studied the variability of
the derived solutions.  We expected to see some variations caused by
effects such as star spots passing over the star's earthward
hemisphere or perhaps micro-pulsations. Model fits would be
`disturbed' by any such non-modelled effects.}

	\item{A more rigorous approach was then taken to explore the determinacy
of the model fits: Markov Chain Monte Carlo (MCMC) methods.  A Markov
Chain is a random process where an object can move from one state to
another with some underlying probability.  We constructed a Markov Chain
to find its stationary distribution, containing the estimates of transit
parameters.  Details on MCMC can be found in Brooks~{\em et al.} (2011).
We performed MCMC on data binned across many planetary orbits (reducing
the noise and hopefully removing short term non-modelled effects),
comparing the derived distributions with the previous results for each
system. { Again, we validated our fits with those from the
literature.} Results from this step were compared with those from earlier
steps to check for consistency, and lend confidence that we had
implemented MCMC correctly.}

	\item{Finally we fitted five systems marked as ``candidates'' in
the NEA, which had no published parameters at the time of the
project.}

\end{itemize}

We were particularly interested in comparison with the algebraic model
of Budding (see Budding {\em et al.}, 2016a, for background on this
methodology) and the parameter `uncertainties' this method provides,
particularly in light of the lower computing loads compared to MCMC
methods.  

A subtle point we note here that the data itself -- regardless of other
physical effects, such as star-spots -- is affected by `Poissonian'
noise relating to the distribution of arrival times of bunches of
photons.  Our approach is to regard all non-modelled effects as `noise',
but the problem is that systematic physical effects are different at
different times, and so treating them as another kind of `white' noise
may detract from the determinacy of the data.  

%-------------------------------------------

\section{Analysis}

\subsection{The model}

{ We decided to build from `first principles' a simple model of how the combined 
light of a single planet orbiting its host star would vary with time. }
A simple light curve model was built using the following parameters: the
ratio of the host star's radius to the planet's orbital radius $r_{1}$, 
the ratio of the planet radius to the stellar
radius $k$, the orbital inclination $i$, the planet's geometric albedo
$A$, and a linear limb darkening coefficient $u$.  { The ratio of the planet's apparent
disk to that of its host star is the key driver of transit depth, while inclination (or the
angle of the planetary orbit relative to our line of sight) indicates the path taken by
the planet across the stellar disk, and is a factor in the duration of the transit. The
geometric albedo was used to model the amount of light from the host star reflected by
the planet.  As the planet orbits about its star, it exhibits phases as `seen' from Earth, varying the
amount of light reflected in our direction and hence the flux of the combined system of star 
and planet (reflection effects).} Two additional
parameters were the phase offset and stellar luminosity.  { Phase 
offset allowed the
phase of the transit to shift in orbital phase to a best fit (a shift on the `time' axis
of the light curve), while the stellar luminosity allowed gross fitting on the `vertical' axis (or
flux)}

The transit formulae of Mandel~\& Agol (2002) were employed, with the projected
distance $d$ between the centres of the stellar and planetary discs
taken from page 258 of Budding~\& Demircan (2007).  The reflection
effect calculations were based on Sobolev (1975) and Charbonneau {\em et
al.} (1999).  We assumed that the transiting planet was small compared
to its host star, and so did not take fully into account the gradient in
the stellar brightening with radius when calculating the stellar
brightness blocked by the planet during transit. We instead assumed that
the obscured light across the portion of the stellar disc obscured by
the planet all had the same brightness as that blocked by the centre of
the planetary disc.  { This assumption is often called the 
``small planet approximation'' in the literature (see, e.g., Nutzman {\em et al.}, 2009).}
Inclination $i$ follows the usual convention
adopted by eclipsing binary light curve models, i.e., when $i =
90^{\circ}$ the planet's orbital plane is in line with our line of sight. 
Eccentric orbits were not included into the model. This, along with
additional limb darkening models and removal of the ``small planet''
approximation, are planned in further work. We refer the reader to the
discussion of Budding {\em et al.\ } (2016a) on the complexity of limb
darkening models and the information content of the modelled data, and
also to the work below in this paper where a quadratic limb darkening
model was included into fitting.

%New summary table below

\begin{table*}[htpb]
\begin{center}
\begin{tabular}{cccccccc}
\hline
System & Parameters                 &   nlsLM  &     GA  &     PSO &     SA     &       ILOT                     &       NEA \\
\hline\hline
Kepler 1b & $k$& 0.125     & 0.125  & 0.126   & 0.125     & 0.1275 $\pm$ 0.0007 & 0.12539 \\
...             & $r_{1}$      & 0.123     & 0.122  & 0.123   & 0.123     & 0.122 $\pm$ 0.002     & 0.126 \\
...             & $u$                         &    -          & 0.584  & 0.587   & 0.590     & -                                  & 0.56 \\
...             & $i$                          & 84.086  & 84.113  & 84.111 & 84.105   & 84.3 $\pm$ 0.2            & 83.872 \\
Kepler 2b & $k$	& 0.077      & 0.078    & 0.077    & 0.078      & 0.0761 $\pm$ 0.0002    & 0.077524 \\
...             & $r_{1}$      & 0.235      & 0.238    & 0.236    & 0.236      & 0.2155 $\pm$ 0.0003     & 0.2407 \\
...             & $u$                         &    -           & -            & -            & 0.456      & -                                     & 0.51 \\
...             & $i$                          & 83.928    & 83.721   & 83.917  & 83.641    & 87.2 $\pm$ 0.5              & 83.14\\
Kepler 8b &$k$		& 0.096      & 0.095    & 0.097    & 0.097      & 0.0972 $\pm$ 0.0008     & 0.095751 \\
...             & $r_{1}$      & 0.148      & 0.144    & 0.149    & 0.149      & 0.146  $\pm$ 0.002       & 0.1459 \\
...             & $u$                         &    -           & -            & 0.527    & 0.512      & -                                     & 0.52 \\
...             & $i$                          & 83.730    & 84.040   & 83.712  & 83.704    & 84.0 $\pm$ 0.4              & 83.978\\
Kepler 77b & $k$	& 0.101      & 0.099     & 0.101    & 0.102      & 0.1081 $\pm$ 0.0002      & 0.099241 \\
...             & $r_{1}$      & 0.114      & 0.107     & 0.113    & 0.149      & 0.147 $\pm$ 0.008      & 0.1024\\
...             & $u$                         &    -           & -            & 0.556    & 0.512      & 0.600 & 0.505 \\
...             & $i$                          & 86.459    & 87.306   & 86.651  & 83.704    & 83.59 $\pm$ 0.06 & 87.998\\
\hline
\end{tabular}
\caption{\label{tab:K1_initial} Estimated physical parameters for Kepler~1b, 2b, 8b, and 77b
based on a single transit (see text) for each system. $i$ is in units of
degrees.  ILOT is the Information Limit Optimisation Technique of
Budding, nlsLM is the Levenberq-Marquardt method, GA the
genetic algorithm, PSO particle swarm optimisation, SA simulated
annealing, and NEA are the NASA Exoplanet Archive published values. In
the absence of error estimates we give our results to three decimal
places}
\end{center}
\end{table*}

\subsection{Initial tests}

A number of tests were made of the model.  The first set concerned
whether we could recover the input parameters of the model using
standard optimisation techniques.  The model was used to generate
simulated light curves, to which Gaussian noise was applied (at the 3\%
to 0.0003\% levels in powers of 10). { This noise is defined 
as the standard deviation of the flux due to the host star by itself (i.e., outside
transits and not including interaction effects such as the reflection effect).}
We used the Levenberg-Marquardt
algorithm for these initial tests, which is a combination of gradient
descent and the Gauss-Newton algorithms. This is implemented as the
$nlsLM$ function in the R $minpack.lm$ package{ (Elzhov {\em
et al.}, 2016)}.  The R statistical
programming environment (R Core Team, 2014) was used throughout this
project { (R is available from https://www.r-project.org)}.
No successful solutions were found at the 3\% noise level, but
by 0.3\% solutions were being found.  By the time noise was at 0.03\%
the derived parameters for the radii, orbital distance, and inclination
were effectively the same the input values. Limb darkening had to be
fixed in most runs, as $nlsLM$ was not able to settle on reasonable
estimates even for linear limb darkening co-efficients.  The published
noise level for the Kepler data is of the order 0.03\% (see later for
noise levels in the modelled data), lending confidence that we would be
able derive appropriate estimates.  Of course, this is a `rule of thumb'
as the relative scatter in the light curves will be different for
different systems depending on their brightness (stellar magnitude).

%----New Summary table below

\begin{table*}[htpb]
\begin{center}
\begin{tabular}{ccccc}
\hline 
System      & Variable                        & Median        & 95\% Interval            & ILOT   mean $\pm$ 4 std.err.   \\
\hline\hline
Kepler~1b   & $k$     &  0.125         & (0.124, 0.125)            & (0.125, 0.130) \\
 ...                & $r_{1}$          &  0.123         & (0.122, 0.124)            & (0.114, 0.130)  \\
 ...                & $i$                               & 84.075       & (84.015, 84.142)         & (83.5, 85.1)  \\
 Kepler~2b  & $k$     & 0.077          & (0.077, 0.078)            & 0.075, 0.077) \\
 ...                & $r_{1}$          & 0.235          & (0.233, 0.245)            & (0.214, 0.217) \\
 ...                & $i$                               & 83.835        & (83.415, 84.668)        &  (85.2, 89.2)  \\
 Kepler~8b  & $k$     & 0.096          & (0.093, 0.104)            & (0.094, 0.100)   \\
 ...                & $r_{1}$          & 0.141          & (0.131, 0.176)            & (0.138, 0.154)  \\
 ...                & $i$                               & 84.384        & (83.512, 86.793)       & (82.4, 85.6) \\
 Kepler~77b & $k$     & 0.099         & (0.097, 0.103)            & (0.100, 0.116) \\
 ...                &$r_{1}$           & 0.101          & (0.097, 0.123)             & (0.100, 0.116) \\
 ...                & $i$                               & 88.064       & (86.559, 90.000)         & (83.35, 83.83) \\
 \hline
\end{tabular}
\caption{\label{tab:summary_multiple} Summary of estimated parameters of 30 light curves for Kepler~1b, 2b, 8b, and 77b. 
To ease comparison of the results, a column is given providing the ILOT upper and lower estimates for $4 \sigma$
from the previous tables for the systems is given. $i$ is in degrees}
\end{center}
\end{table*}

Having established that we could recover the input parameters from data
sets based on the model combined with noise, we moved on to fit NEA data
for the well-known cases of Kepler~1b { TrES-2b} (using
Kepler data collection quarter 1), Kepler~2b { HAT-P-7b}
(Qtr 2), Kepler~8b (Qtr 2), and Kepler~77b (Qtr 5).  
{ We used only a subset of the full data available for these systems, 
as we were following an iterative approach to our analysis. No doubt the precision of our modelling could be improved
with more data, and in later papers exploring the {\em Kepler} transit data we intend to do
this. The current paper concerns more methodological research, rather than intensive studies of particular examples.
In a similar way, Rhodes \& Budding (2014) used only single quarter data-sets
(on 16 systems) to test their {\sc winfitter} technology.
}

Our object was to
see if parameter estimates from our best-fit model would be in line with
the published parameters.  Short cadence data sets were folded using the
periods published on the NEA.  Data were binned in time by 0.0002 days
for transit ingress and egress, and 0.005 outside these times.  30
consecutive orbits were processed for each system, with each transit
being individually modelled.  Orbital period estimates were made via
linear regression on the transit midpoints. Standard errors (in
$10^{-6}$ days) were 3, 7, 7, 4, for Kepler 1b, 2b, 8b, and 77b
respectively.  The NEA published periods fell within the 95\% confidence
intervals of the fits, so we were comfortable with these NEA estimates,
and continued our use of them.

We also wanted to explore choice of optimisation technique, as $nlsLM$
has been reported sensitive to starting values, to the point that
convergence might not be reached (see Ford, 2005).  We tested the
simulated annealing (Betsimas~\& Tsitsiklis, 1993), genetic algorithm
(see Mitchell, 1998), and particle swarm methods (see Rini {\em et al.},
2011) as implemented in the $GenSA$ { (Xiang {\em et al.},
2013)}, $genalg$ (function $rbga$; Cortez, 2014), and $psoptim$
{ (Ciupke, 2016)} R libraries.  A modelling technique
originally developed by Budding was also tested --- the Information
Limit Optimisation Technique (ILOT; Banks~\& Budding, 1990).  ILOT has
several interesting points:

\begin{itemize}

\item The relatively simple and compact algebraic form of the fitting
function, which allows large regions of the $\chi^2$ parameter space to
be explored at low computing cost.

\item The $\chi^2$ Hessian (see, for example, Bevington, 1969) can be
simply evaluated in the vicinity of the derived minimum.  Inspecting
this matrix, and in particular its eigenvalues and eigenvectors, gives
valuables insights into parameter determinacy and interdependence.

\item The Hessian can be inverted to yield an error matrix.  This must
be positive definite if a determinate, ``unique" optimal solution is to
be evaluated.  ILOT considers strict application of this provision
essential to avoiding over-fitting the data.

\end{itemize}

ILOT uses three different optimisation techniques:
\begin{enumerate}
\item Parabolic interpolation for single parameters, in a step by step mode;
\item Parabolic in conjugate directions (``Powel''); and
\item Vector (in all parameters) mode
\end{enumerate}

The program switches between these modes depending on the convergence
rate and user defined limits.  ILOT has been applied to a variety of
problems, ranging from radial velocity curves for binary stars (Banks
{\em et al.}, 1990), through chromospherically active stars and
interacting binaries (e.g., Zeilik {\em et al.} (1988), Zeilik {\em et
al.} (1994), to orientations of spiral galaxies (Banks {\em et al.},
1994) and radial luminosity functions (Banks {\em et. al.}, 1995). 
The technique has been applied to the light curves of exoplanets in the
implementation called {\sc winfitter} (see., e.g., Budding {\em et al.},
2016a and 2016b), which we make use of in this paper.

We were particularly interested to see if the parameter uncertainties
provided by ILOT were reasonable estimates.  Published ILOT results
(Budding {\em et al.}, 2016a) were available for three of the test
systems, and we ran ILOT fits for Kepler-77b in this paper.  The data
for 77b (used for the ILOT fitting) consisted of the entire available
short cadence data from {\em Kepler} (quarters 1 to 17 inclusive),
binned down to a thousand uniformly spaced points (around one orbital
cycle).

%-------- Master table for Rhat -----------------

\begin{table*}[htpb]
\begin{center}
\begin{tabular}{cccccccc}
\hline 
  System              & $k$  & $r_{1}$  & $u$  & $cos(i)$ & $offset$ & $U$ & $\sigma$ \\
\hline\hline
Kepler 1b   & 1.0061   & 1.0055  & 1.0064 &  1.0058 & 1  & 1 & 1 \\
Kepler 2b  & 1.0023   & 1.0017  & 1.0006 &  1.0016 & 1  & 1 & 1 \\
Kepler 8b  & 1.0062   & 1.0038  & 1.0153 &  1.0031 & 1  & 1 & 1 \\
Kepler 77b  & 1.0001   & 1.0004  & 1.0002 &  1.0004 & 1  & 1 & 1 \\
KOI 760.01  &1.001789  &1.003602  &1.004035  &1.004232  &1  &1  &1 \\
KOI 767.01   &1.010862  &1.009553  &1.003716  &1.010511  &1  &1  &1 \\
KOI 802.01  &1.001221  &1.003294  &1.003736  &1.003597  &1  &1.000292  &1 \\
KOI 824.01  &1.032533  &1.026074  &1.005429  &1.040743  &1  &1.000518  &1\\
\hline
\end{tabular}
\caption{\label{tab:rhat_table} $\hat{R}$ of transit parameters for the systems modelled in this paper}
\end{center}
\end{table*}

We first compared results of the different optimisation methods.  Table~\ref{tab:K1_initial}
lists the results of these fits to one transit (the first in each
system's selected dataset).   Unless otherwise noted, only the light
curve around the transit was modelled in this paper.  The best fits for
all the optimisation techniques are similar for Kepler 1b.  Limb
darkening was not an optimised parameter in the ILOT fits, and we noted
above that $nlsLM$ struggled when limb darkening was included into the
fitted parameter set.  For Kepler~2b, the ILOT results appeared to be
the outlier, with the other techniques agreeing.  Results were in
general agreement across the remaining two systems.

We concluded that our model, although simple, was producing parameter
estimates in line with reasonable expectation.  The close agreement of
the different optimisation techniques appeared to indicate a simple,
fully convex search space around the optima with no significant local
minima acting as 'traps' during the searches.  Having gained confidence
that we could model real transit data, and derive optimal parameters in
line with those published on NEA, we moved on to fitting the remaining
transits in our data sets with the intention of examining any scatter in
the derived parameters across these fits.

We then continued fitting across the remaining 29 transits in the data
sets for each system, using just the $nlsLM$ method as we had shown it
was equivalent to the others in the tests above.  Computing run times
with $nlsLM$ were shorter than for the other methods, and our testing
above gave us confidence that it was deriving optimal parameters in line
with those from more CPU-itensive methods.  Given the issues fitting
$u$, we kept this as a fixed parameter (using the NEA values). The
optimisation methods that had provided estimates of $u$ gave results in
essential agreement with the NEA estimate, so we felt the compromise of
using $nlsLM$ and fixing $u$ was acceptable for this preliminary look
into the scatter of the derived parameters. At this stage it was  not
really clear if $u$  could be estimated from the fitting or not, even for
the simple linear model. However, `ILOT' philosophy would point to more
determinacy for the parameter set if a reasonable value of $u$ is first
selected and that value retained in subsequent fittings.

Table~\ref{tab:summary_multiple} give the results of these fits.  In
order to compare the interval estimates, we applied Chebyshev's
Inequality to generate ILOT interval estimates.  According to
Chebyshev's Inequality, at least $\frac{15}{16}$ of the values of any
distribution are within four standard deviations of the mean.  We use
mean $\pm$ four standard deviations as the interval estimates of ILOT. 
The ILOT intervals for Kepler 1b are wider than the 95\% quantile
estimates from the $nlsLM$ fits.  None of the 95\% quantile estimates
for { Kepler-2b} over-lapped with the ILOT intervals, while
it was a mix for the remaining two systems.

There was clearly a wide spread in the parameter estimates for the
different transit fits, which could be due to additional effects
mentioned above ``disturbing'' the light curve.  Folding and binning
across multiple orbits should reduce the impact of noise and effects
such as quickly evolving spots.    
{
Folding and binning across multiple orbits
should reduce the impact of white noise, as they have
zero mean and are uncorrelated. However, removing
the impact of correlated noise, i.e. `red noise', is more
difficult. In some cases, where the red noise has a period in harmony with the folding period,
binning may reinforce the red noise and lead to biasing.  Some
research teams (e.g., Pont~{\em et al.}, 2006) have conducted studies in this area and
have come up with techniques to eliminate the non-white noise from data, for example, a wavelet-based
whitening filter (see Tenenbaum~{\em et al.}, 2010). No doubt
reducing red noise could improve the precision of fitted
parameters, but as our initial test results were in line
with those published on NEA, we did not concentrate
on this and instead proceeded on to the Markov Chain Monte
Carlo (MCMC) method.
}

\subsection{MCMC fits}

{ MCMC allows us to construct a Markov process such that the stationary distribution is the same as
our target distribution.  Random samples are generated from the process.  After a sufficient number of samples, and with
the influence of data, the chain becomes close enough to the stationary distribution and thus the samples give a good approximation of
the target distribution.  This is known as the convergence of MCMC (see Sinharay, 2003).  

Several approaches are applied to assess the convergence, including trace plot diagnostic,
auto-correlation function (ACF) plots diagnostic and Gelman-Rubin convergence diagnostic ($\hat{R}$). 
Trace plot diagnostics are an intuitive approach.  We compared the trace plots of different chains to
see if they overlapped with each other.  ACF plots present the correlations among samples and help to indirectly check the convergence of MCMC.  For chains with ACF decaying fast (as lag increases), it takes shorter time for them to explore
the whole sampling space (see Sinharay, 2003)). 

The Gelman-Rubin (Gelman~\& Rubin, 1992) convergence diagnostic ($\hat{R}$), also known as the Potential Scale Reduction Factor,
is defined to be the square root of ratio of posterior variance over within-sequence variance of the Markov chains,
taking sampling variability into account.  Within-sequence variance refers to the variance of random samples
within each Markov chain and between-sequence variance is the variance among different chains. 
Putting it another way, the diagnostic evaluates MCMC convergence by comparing the estimated between-chains and within-chain variances for each model parameter, across multiple chains being run on the problem. Large differences between these variances indicate nonconvergence. 
Posterior variance takes both within-sequence variance and between-sequence variance into account. 
MCMC is considered converged when $\hat{R}$ is approximately 1.  As discussed in Brooks~\& Gelman (1997),
if $\hat{R}$ is less than 1.2 the chains are approximately converged (see also section 7.3 of Givens~\&
Hoeting, 2013, and references within).
}

Similar to the work in sections above, a stepwise approach was taken.  First we
tested two MCMC methods: the RWMH (Random Walk Metropolis Hastings) and
HMC (Hamiltonian Monte Carlo) methods.  We used Kepler 93b in this test
since Ballard {\em et al.} (2014) has also applied a MCMC method on this
system.  We could compare our results with those of the Ballard team. If
our results were comparable to that paper's, we could feel more
confident to apply our implementation to other systems.  We would then
apply the method to the systems modelled above, and if in good
agreement, move on to model systems that had no published parameters.

The RWMH was implemented using the function $rwmetrop$ from the
$LearnBayes$ R library { (Albert, 2014)}.   For a Markov
Chain to converge, it has to find the stationary distribution and
explore it rapidly.  The RWMH algorithm is able to find the
distribution, but it can take a long time to achieve full convergence
due to incoherent exploration.  HMC is a faster method, as it explores
the posterior distribution in a coherent manner. It is more efficient
that RWMH when dealing with high dimensional problems with correlated
variables (see Wang, Mohamed \& de Freitas, 2013).  HMC is implemented
by the $rstan$ R package { (Stan Development Team, 2016).}

The ACF (auto-correlation function) plots of the RWMH estimates
demonstrated much higher dependence than the HMC samples, indicating that
the chains may take a longer time to achieve full convergence.  The
runtime for RWMH was approximately double that for HMC.  These results
suggest that HMC has better performance in this case than RWMH for our
current case.

We compared the pairwise plots and histograms generated from the RWMH
and HMC runs with those in Ballard {\em et al.}  The HMC results were
more similar to Ballard {\em et al.} than those from RWMH.  Considering
all these points, we decided to use the HMC method for the subsequent
work.

\subsubsection{Known Systems}

\begin{itemize}

\item Kepler 1b: the autocorrelations for this system decay to zero
quickly, suggesting that the samples are approximately independent.  The
trace plots demonstrate good mixing and that the chains are exploring
the posterior distribution rapidly.  All four ACFs decay to null quickly
and the trace plots agree with each other, therefore we believe that the
chains have converged.  All parameters have unimodal distributions (see
Figure~\ref{fig:3}).  Figure~\ref{fig:3} was presented in Budding {\em et al.} (2016b). 
Now we have applied this method to more system.  In passing, we note that the correlation plots for all other 
systems looked very similar to this chart, so in the sake of space we have not
reproduced the charts in this paper, instead, interested parties may contact the
authors for these additional charts.  The correlations plots suggest that
$k$ , $r_{1}$ , $u$, and $cos(i)$ have
higher correlations than the others.  For all four systems, all $\hat{R}$ statistics were
close to 1 (see Table~\ref{tab:rhat_table} for convergence statistics and
Table~\ref{tab:K1_hcm} for the parameter results).  

% Pairwise correlation diagram
\begin{figure*}[hbtp]
\center{\includegraphics[width=0.7\textwidth]{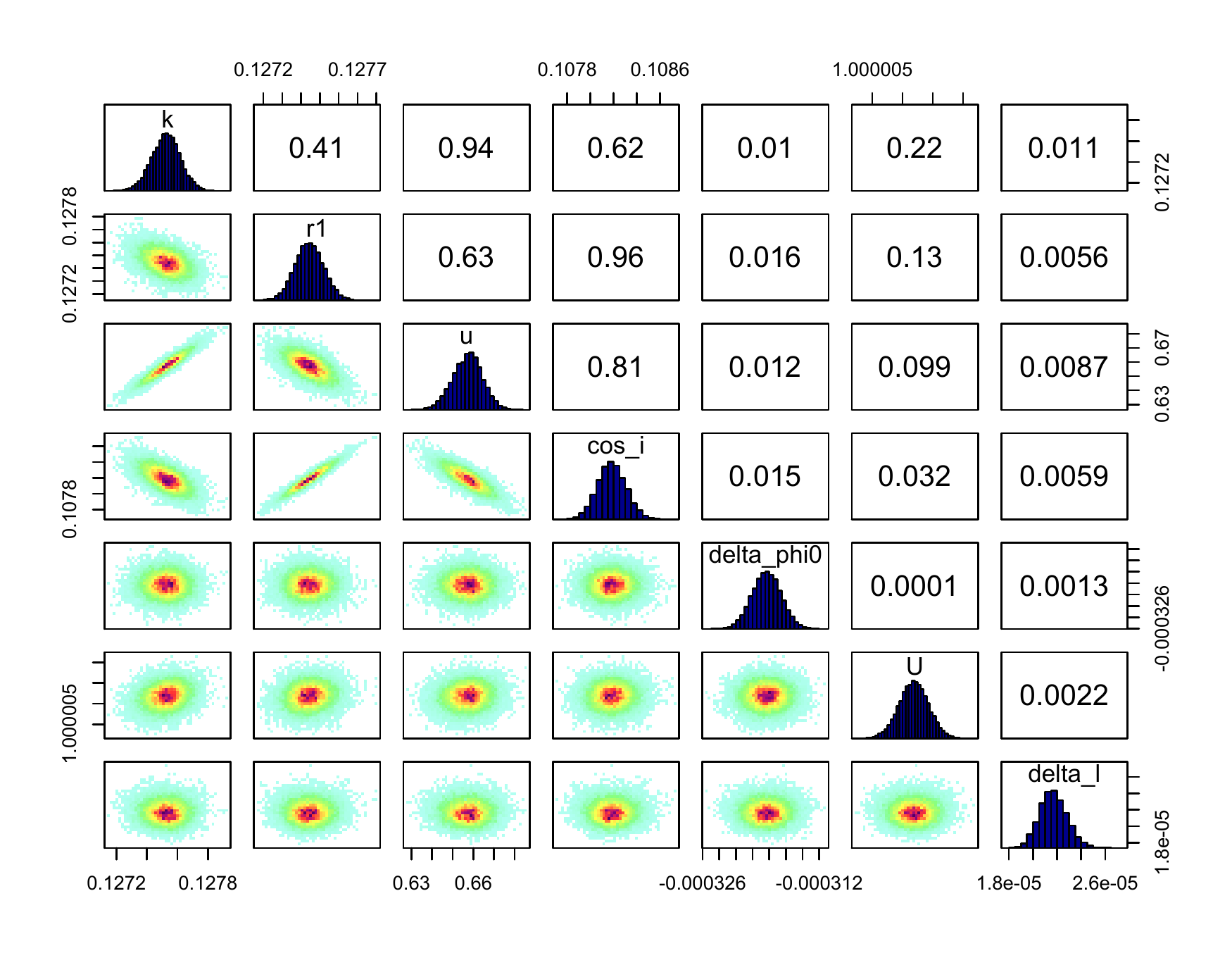}}
\caption{Pairwise Correlation Diagram of 7 estimated parameters of 
{ Kepler-1b.  The columns correspond to each of the model parameters, as 
discussed in Section~2.1 (``The model"), e.g., $k$ is the ratio of the planet radius to the
stellar radius, and so on.  The pairwise correlation coefficients are in the upper right boxes, e.g.,
0.41 for between $r_{1}$ and $k$.  The histograms are the distributions for each of the parameters
The lower left (below the diagonal) plots the joint distribution between two parameters at a time, 
giving an insight into the n-dimensional surface being optimised over.  The colours correspond
to frequency, or the number of times the MCMC was found in this point.  Red corresponds to
most frequent, and is found in the centre of the distributions, with counts falling away from these
points.
}}
\label{fig:3} % reference
\end{figure*}

\item Kepler 2b: The histograms show that the distributions of all
parameters are unimodal.  The pairwise correlations plots  show that
$r_{1}$  and $cos(i)$ have the strongest relationship, while
the others are less related.  All $\hat{R}$ statistics were close to 1.

\item Kepler 8b: Again, all parameters have unimodal distributions.   The correlations
plots suggest that the parameters do not have linear relationships. 
Convergence was good (see Table~\ref{tab:rhat_table}), but the 95\%
quantiles are wide, indicating a weak solution.

\item Kepler 77b: is similar to Kepler 8b, with wide quantiles but good
convergence statistics.  { Gandolfi {\em et al.} (2013) present a more
detailed MCMC-based analysis than this paper's, including radial velocity measurements
with {\em Kepler} photometry.  We note that our paper's error estimates are an order
of magnitude larger than those of Gandolfi {\em et al.}, for instance they estimate
inclination as $88.00 \pm 0.11$ degrees and the planet to star radius as $0.09924 \pm
0.00026$, with error bars being defined to be at the 68\% confidence limits.  We plan
to revisit Kepler 77b with a more sophisticated model (as discussed above) and see if
an improved model leads to a reduction in our error estimates and towards those of
Gandolfi {\em et al.}}

\end{itemize}

A feature of the HMC fits is just how large the 95\% uncertainties from
the fitting are, reflecting the difficult nature of `untangling' the
various effects we have modelled. The HMC solutions are generally
consistent with the model fits before, but now have estimates of the
parameter reliability, plus we were able to include $u$ as a free
parameter more confidently.  Not all the HMC intervals overlap with the
intervals constructed from the multiple light curves analyses.  The ILOT
`errors' for Kepler-1 were similar to the uncertainties from HMC.  The
HMC result for Kepler-2 was quite different from ILOT, just as ILOT's
solution had been different to those of the other optimisation methods
tested at the beginning of this study.  HMC also struggled with the
Kepler-8 data, giving large uncertainties, as with Kepler-77.  We
discuss the effect of lowering signal to noise later in the paper. There
is also the issue of `chi-squared valleys' or `local minima' --- about
which some optimization methods `handle' better than others, not getting
caught in these secondary minima.  MCMC methods are particularly good at
such escapes, although as noted above we do not appear to have
over-parameterised the transit light curve data and therefore the
solution search space appears to be highly convex.  The MCMC correlation
diagrams confirm this interpretation, along with the not surprising
correlations between the three key parameters of planetary radii
($r_{1}$), ratio of the planetary and stellar
radii ($k$) and the limb darkening ($u$).  These relations with limb darkening
somewhat complicated comparisons with the simpler modelling above, and
speaks towards the necessity of the more computing intensive method of
MCMC to properly explore the search space.  As we will see with later
systems in this paper, the optima can be more complicated than we
expected at this stage of the study.  Offset parameters (in phase and
flux) were clearly Gaussian, as was the noise (`delta\_l' in the
correlation charts). 

The results for Kepler~8 are quite different to the initial estimates,
but within the estimated uncertainties.  The difference could be due to
data treatment (e.g., data folding and binning) as well as the outliers
in the multiple light curve results.  However, we believe that the HMC
estimates are more accurate, given that we only tested 30 light curves
in the individual fits above.  

% New Summary table follows
\begin{table*}[htpb]
\begin{center}
\begin{tabular}{cccc}
\hline 
System & Parameters                 & Point Estimate       & 95\% Interval Estimate \\
\hline\hline
Kepler 1b  & $k$    & 0.126                      & (0.122, 0.128) \\ 
...            & $r_{1}$          & 0.121                      & (0.120, 0.122) \\
...            & $u$                             & 0.638                      & (0.555, 0.711) \\
...            & $i$                              & 84.238                    & (84.110, 84.355) \\       
...            & $\sigma$                    & 0.000057                & (0.000048, 0.000070) \\
Kepler 2b & $k$    & 0.0777                     & (0.0775, 0.0779) \\ 
...            & $r_{1}$          & 0.236                       & (0.235, 0.238) \\
...            & $u$                             & 0.461                     & (0.451, 0.472)\\
...            & $i$                              & 83.545                    & (83.293, 83.304) \\       
...            & $\sigma$                    & 0.000041                & (0.000035, 0.000047) \\
Kepler 8b & $k$    & 0.094                    & (0.090, 0.097) \\ 
...            &  $r_{1}$          & 0.138                     & (0.124, 0.156) \\
...            & $u$                             & 0.658                     & (0.399, 0.811)\\
...            & $i$                              & 84.759                    & (83.082, 86.244) \\       
...            & $\sigma$                    & 0.00044                & (0.000037, 0.000052) \\
Kepler 77b & $k$    & 0.099                     & (0.096, 0.104) \\ 
...            & $r_{1}$          & 0.112                     & (0.100, 0.123) \\
...            & $u$                             & 0.558                     & (0.378, 0.721)\\
...            & $i$                              & 86.863                    & (85.585, 88.890) \\       
...            & $\sigma$                    & 0.00073                & (0.00061, 0.00088) \\
\hline
\end{tabular}
\caption{\label{tab:K1_hcm} Summary of estimated parameters from HMC for Kepler 1b, 2b, 8b, and 77b}
\end{center}
\end{table*}

\begin{figure*}[htbp]
\center{\includegraphics[width=0.7\textwidth]{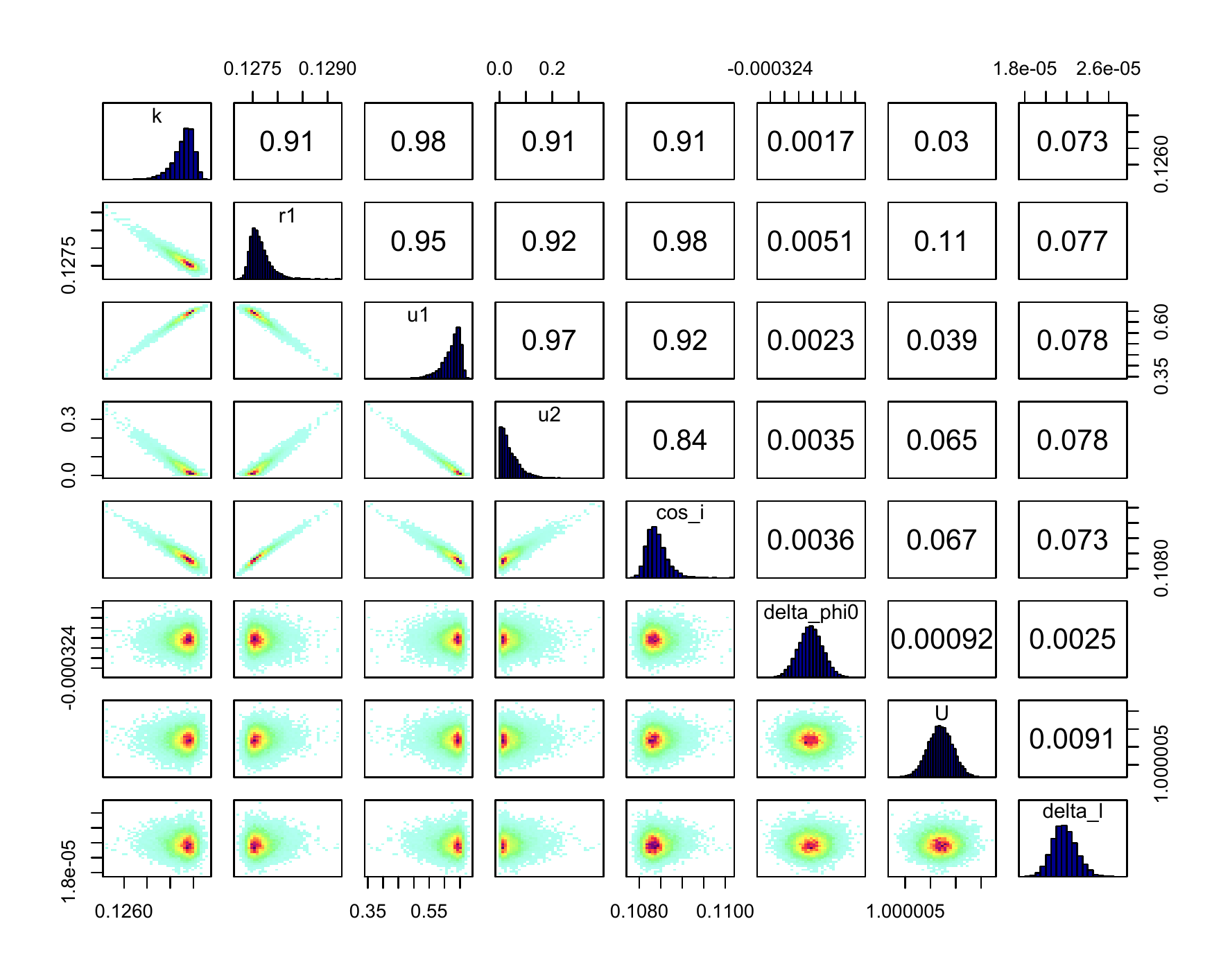}}
\caption{\label{fig:aaa} Pairwise Correlation Diagram of 8 estimated parameters of Kepler 1b. $u_1$ and $u_2$ are parameters
for the quadratic limb darkening model}
\end{figure*}

We considered whether to add more detailed modelling of the limb
darkening.  A test was therefore made using a quadratic limb darkening
law for Kepler~1b. The limb darkening coefficients were not well defined
(see Table~\ref{tab:K1_quadratic_rhat} and \ref{tab:K1qhcm}, plus
Figure~\ref{fig:aaa}).  We saw no improvement in going to this model,
and so remained with a simple linear model for the remaining fits.
Within ILOT philosophy, including more terms in the l.d.\ expression
dilutes the determinacy of individual parameters, even when it does not
create a general breakdown of determinacy due to the failure of the
Hessian to remain positive definite. And with that, the higher order
terms in that expression are expected to be small, or very small, in
comparison to their errors.

%---------------------------------
\begin{table*}[htpb]
\begin{center}
\begin{tabular}{ccccccccc}
\hline 
                & $k$  & $r_{1}$  & $u_1$ & $u_2$  & $cos(i)$ & $offset$ & $U$ & $\sigma$ \\
\hline\hline
$\hat{R}$  &1.037882 &1.040101 &1.048493 &1.052874 &1.030407 &1 &1 &1\\
\hline
\end{tabular}
\caption{\label{tab:K1_quadratic_rhat} $\hat{R}$ of transit parameters for Kepler 1b with quadratic limb darkening model}
\end{center}
\end{table*}

\begin{table*}[htpb]
\begin{center}
\begin{tabular}{ccc}
\hline 
Parameters                 & Point Estimate       & 95\% Interval Estimate \\
\hline\hline
$k$    & 0.1273		 & (0.1267, 0.1276) \\ 
$r_{1}$          & 0.1276		& (0.1274, 0.1281) \\
$u_1$			&0.6220		&(0.5394, 0.6612)\\
$u_2$                         & 0.0420		& (0.0016, 0.1401) \\
$i$                              & 83.776	          & (83.745, 83.797) \\       
$\sigma$                    &0.00002		 & (0.00002, 0.00002) \\
\hline
\end{tabular}
\caption{\label{tab:K1qhcm} Summary of estimated parameters from HMC for Kepler 1b with quadratic limb darkening model}
\end{center}
\end{table*}

\subsubsection{ {Candidate Systems}}

The HMC fits on the known systems gave comparable results to the
literature and the previous fits in this study.  We were therefore
confident to apply the model to four `new' systems without published
parameters.  These are KOI 760.01, 767.01, 802.01, and 824.01. We
adopted the NEA orbital periods after confirming that the NEA-assigned
periods were reasonable via linear regressions (see
Table~\ref{tab:New_Period}).

The light curves are not as `clean' as the previous systems modelled.
KOI 802.01 and 824.01 have relatively long periods, reducing the number
of transits available for modelling.  There were only 5 transits
available for 760.01, 30 for 767.01, 11 for 802.01, and 9 for 824.01. 
We did model these transits separately, but the `error bars' were large
(even $r_{1}$ for 767.01 had a 95\% interval range of 0.103 to
0.222, so we decided to go straight through to MCMC modelling.

Before discussing the results, we will briefly consider signal to noise.
If we compare transit depth to the noise $\sigma$ it should be no
surprise that the S/N (signal to noise) levels are lower for the
{ candidate systems}.  Kepler 1b has a signal to noise ratio of about 252,
2b 164, 8b 22, and 77b 16.  By comparison, the S/N for KOI 760.01 is 7,
for 767.01 19, 802.01 2.3 , and 824.01 7.  We therefore expected wider
`uncertainties' in our results for the { candidate systems}, and that our
parameters would be best defined for KOI 767.01 of all the { candidate
systems}.

\begin{itemize}

\item{KOI 760.01: Table \ref{tab:uK760_hcm} gives the results of these
fits, and for the other three { candidate systems}.  
The limb darkening is approximately uniformly distributed,
indicating poor definition.  Other
parameters have unimodal distributions.  Only $r_{1}$ and
$cos(i)$ have a strong positive correlation.  Convergence statistics
look good (Table~\ref{tab:rhat_table}) and the chains appear converged. 
 Limb darkening does not seem to be an optimisable variable for this
data set.}

\item{KOI 767.01:  All
parameters are unimodal bar $cos(i)$, which is apparently uniformly
distributed.  Chains appear well converged.}

\item{KOI 802.01: 
$r_{1}$ and $u$ can take any values between 0 and 1,
indicating that we can not optimise these parameters together with the
others.  This system has a low signal to noise ratio, so it is not
unexpected that modelling will be more problematic.}

\item{KOI 824.01:   Similar to KOI
802.01, this system has a low signal to noise ratio.  Parameter $u$ can
take any values between 0 and 1, indicating that we can not optimise it
together with the others.}

\end{itemize}

% Linear Regression for Period Estimation: Table
\begin{table*}[htpb]
\begin{center}
\begin{tabular}{ccccc}
\hline 
System                 & Point Estimate       & Standard Error	  & 95\% Interval Estimate         & NEA \\
\hline\hline
KOI 760.01           & 4.9597                    & 0.0005                     &(4.9577, 4.9617)                    &4.95932 \\ 
KOI 767.01	    & 2.81659                  & 0.00009                   &(2.8164, 2.8168)                    &2.81650 \\
KOI 802.01           & 19.60                      & 0.01                         &(19.569, 19.627)                    &19.62035 \\
KOI 824.01           & 15.375                    & 0.002                       &(15.368, 15.382)                    &15.37563 \\ 
\hline
\end{tabular}
\caption{\label{tab:New_Period} Summary of estimated periods (in days) for 4 `new' systems}
\end{center}
\end{table*}

\begin{table*}[htpb]
\begin{center}
\begin{tabular}{cccc}
\hline 
System & Parameters                 & Point Estimate       & 95\% Interval Estimate \\
\hline\hline
Kepler 760.01 &  $k$    & 0.1066			 & (0.1001, 0.1120) \\ 
...                     & $r_{1}$          & 0.0847 		 & (0.7590, 0.0925) \\
...                     &$u$                             & 0.722 			 & (0.0894, 0.9896)\\
...                     &$i$                              & 85.888                   & (85.279, 86.600)\\       
...                     &$\sigma$                    & 0.0008 			 & (0.0007, 0.0009) \\
Kepler 767.01 & $k$    & 0.1238			 & (0.1226, 0.1250) \\ 
...                     &$r_{1}$          & 0.1176		 	 & (0.1152, 0.1199) \\
...                     &$u$                             & 0.5234			 & (0.4724, 0.5704)\\
...                     &$i$                              & 86.618	                  & (86.329, 86.921)\\       
...                     &$\sigma$                    & 0.0005 			 & (0.0004, 0.0006) \\
Kepler 802.01 & $k$    & 0.1477			 & (0.1456, 0.1499) \\ 
...                     &$r_{1}$          & 0.0163		 	 & (0.0156, 0.0169) \\
...                     &$u$                             & 0.1478			 & (0.0056, 0.4043) \\
...                     &$i$                              & 89.296	                  & (89.247 ,89.352) \\       
...                     &$\sigma$                    & 0.0062			 & (0.0060, 0.0064) \\ 
Kepler 824.01 & $k$    & 0.1214			 & (0.1130, 0.1265) \\ 
...                     &$r_{1}$          & 0.0317		 	 & (0.0256, 0.0363) \\
...                     &$u$                             & 0.4517			 & (0.0428, 0.8618) \\
...                     &$i$                              & 88.671	                  & (88.299, 89.274) \\       
...                     &$\sigma$                    & 0.0032			 & (0.0030, 0.0033) \\
\hline
\end{tabular}
\caption{\label{tab:uK760_hcm} Summary of estimated parameters from HMC for KOI 760.01, 767.01, 
802.01, and 824.01}
\end{center}
\end{table*}

Among the 8 systems, Kepler 1, 2, 8, 77 and KOI 767.01 have S/N above
15.  Their corresponding estimates of limb darkening coefficient $u$
have higher determinacy than the others, i.e., the error bars are
narrower.  Therefore, we have estimated the borderline S/N for limb
darkening coefficient to be 15.  For noisy data with low S/N, it is
preferable to take the limb darkening coefficient derived from
appropriate stellar atmosphere models rather than derive it from the
data. 

{ We note Csizmadia {\em et al.} (2013) made a detailed study of the
determinability of limb darkening coefficients. This study created 2000
light curves using the subroutines of Mandel~\& Agol (2002), assuming
a circular orbit and adding levels of Gaussian noise to give different
levels of signal to noise (S/N).  A genetic optimisation method was then applied
to these synthetic light curves, fitting a transit model.  Figure~3 of their
paper plots the relative uncertainties in derived parameters, including limb 
darkening, against S/N. They note that a S/N of at least 25 and 6 is required
to fit quadratic limb darkening coefficients (i.e., the determinability of the
limb darkening coefficients depends on the transit depth relative to the noise).
From this figure, it can be seen that the limb darkening co-efficients can
be derived subject to an approximately 6\% precision if the S/N ratio is 15, in
line with the current paper and confirming their results.}

We have re-fitted KOI 802.01 (with the lowest S/N)
data using 6 parameters, excluding limb darkening coefficient $u$. As
for $u$, we have fixed a theoretical value 0.41 determined by {\sc winfitter}
program. See tables~\ref{tab:uK802_hcm_nou} and
\ref{tab:uK802_rhat_nou}.

All six transit parameters for KOI 802.01 have uniform distribution and the fitted light
curve looks reasonable. According to tables~\ref{tab:uK802_hcm_nou} and
\ref{tab:uK802_rhat_nou}, the 95\% interval estimates are reasonably
small for all six parameters and the $\hat{R}$ are all close to 1,
suggesting the Markov Chain has converged.

%---------------------------------
\begin{table*}[htpb]
\begin{center}
\begin{tabular}{ccc}
\hline 
Parameters                 & Point Estimate       & 95\% Interval Estimate \\
\hline\hline
$k$    & 0.1482			 & (0.1457, 0.1508) \\ 
$r_{1}$          & 0.0158		 	 & (0.0150, 0.0167) \\
$i$                              & 89.353	                  & (89.277 ,89.410) \\       
$\sigma$                    & 0.0062			 & (0.0060, 0.0064) \\
\hline
\end{tabular}
\caption{\label{tab:uK802_hcm_nou} Summary of estimated parameters (excluding $u$) from HMC for KOI 802.01}
\end{center}
\end{table*}

\begin{table*}[htpb]
\begin{center}
\begin{tabular}{cccccccc}
\hline 
                & $k$  & $r_{1}$   & $cos(i)$ & $offset$ & $U$ & $\sigma$ \\
\hline\hline
$\hat{R}$  & 1.000733 & 1.001587 & 1.001534 & 1 & 0.9999253 & 1 \\
\hline
\end{tabular}
\caption{\label{tab:uK802_rhat_nou} $\hat{R}$ of transit parameters for KOI 802.01}
\end{center}
\end{table*}

We referred to the Kepler Candidate Overview page of KOI 802.01 on the
NEA website to check the validity of our fitted results. However, the
information in that catalog seem to be inconsistent. According to the 
``KOI Transit Results'' section, we could derive $R\textsubscript{Star}$
(of $R\textsubscript{Sun}$) by applying Kepler's Third Law. The derived
radius is smaller than 0.5R$_{\odot}$, consistent with our fitted
results. On the other hand, in the Stellar Properties section,
$R\textsubscript{Star}$ (in R$_{\odot}$) is reported as 0.77 to 0.87(!).
Similarly, the effective temperature of the star is given as around 5700
K, which is close to that of the Sun. But for a solar effective
temperature, the derived stellar radius in KOI 802 would be too small
for any normal star. These findings suggest something unusual in the KOI
802.01 identification. It seems possible that the star identified in the
catalog is not the one being eclipsed. Such a scenario might account for
the inconsistency between the radius derived from transit curves and
from the basis of normal stellar structure models.   This question
supports the need for further investigation and additional observational
data.  { We confirm this by noting Batalha (2014), who gives useful information
on the NEA catalogues and explains
that the Kepler Input Catalog (KIC, see Brown {\em et al.}, 2011 for background on this
catalog) provided the stellar properties listed
on the site. These are in turn derived from ground-based photometry and modelling.
Batalha notes that the KIC ``contains known deficiencies and systematic
errors, making it unsuitable for computing accurate planet properties'' and we urge
diligence be taken for stellar parameters case by case, particularly for the fainter
and therefore harder to measure stars (the KIC gives a Kepler magnitude of 15.562 
for KOI 802, which can be compared with the 11.338 for Kepler-1).}

\section{Conclusions}

This paper has described a simple transit model and its application
using a variety of optimisation techniques to fit selected {\em Kepler}
light curves.  The model was first applied to systems with published
parameters, deriving point estimates in good agreement with the
literature.  Provided that the information content of the data is
respected, the point estimates were effectively independent of the
optimisation method indicating a tidy convex optimisation problem.  
{ The following example might make clear our argument against
overfitting:} a sine wave has four meaningful parameters --- amplitude,
offset, frequency and a non-zero center amplitude.   It is possible to
fit additional parameters to a `noisy' sine wave data set, but how to
interpret these parameters?  { Such over-fitting the
data can cloud meaningful analysis, and we believe it is extremely important
for analysts to keep in mind the information content of the data in
analysis, such as through consideration of S/N and its impact on determinability.}

A key question in this study was whether limb darkening could be derived
from the data sets.  We started with a simple linear model, frequently
finding indeterminacy when even linear limb darkening was included as an
optimisable parameter. The situation did not improve when a quadratic
model was applied (we had wondered if the linear model was too much of a
simplification and subsequently leading to poor fits), and the MCMC
optimisations showed limb darkening to be poorly defined from the data. 
However this test was only one case. Later work will involve trialling
different models and investigating whether statistically robust
co-efficients can be derived for {\em Kepler} transits (taking note of
Espinoza~\& Jordan (2015) and Espinoza~\& Jordan (2016)).  The MCMC fits
above clearly showed correlations between limb darkening and a key
variable of interest in exoplanets: the radius of the exoplanet itself. 
We therefore believe that continued modelling and exploration of the
confidence in limb darkening estimates across different models, systems,
and across different periods (for those systems) is an important task. 
Exoplanet transit light curves provide a `laboratory' for exploration of
limb darkening, allowing refinement of our stellar models 
{ (see also Csizmadia {\em et al.} , 2013).} 

Such an investigation will require improvement to the rather simple
model used in this paper, ensuring that we have removed other physics
before trying to fit limb darkening models. We further intend to improve
the fitting model, noting the deficiencies described earlier in the
paper such as use of the small planet approximation, and continue our
investigation into the parameter distributions of model fits to the {\em
Kepler} data. 

Such an improved model will also allow better comparison with results
from {\sc winfitter}, which is a more sophisticated model. The
comparisons between {\sc winfitter} and MCMC in this paper have,
unfortunately, been inconclusive.   We have not really been comparing ``like with like".
As noted above, {\sc winfitter} uses
an optimized chi-square fit of the light curve to the physical
parameters of the system such as mass, luminosity, period and
separation.  It is much faster in computer run times than MCMC, and if
it can be shown that its estimates of the accuracy of parameters are
consistent with MCMC, then it would be a useful tool for parameterising
transit light curves, particularly at large scale (i.e., an automated
system running across large data sets of multiple systems). {\sc
winfitter} includes the relevant proximity effects such as radiative
interaction (reflection effects), tidal and rotational distortions,
gravity brightening (ellipticity effects), limb darkening, Doppler
beaming effect, and orbital eccentricity (if present).  A further
advantage of {\sc winfitter} is that it also produces a distortion-wave
light curve that is the difference between the observed and model light
curves. This `difference curve' can then be analyzed by {\sc
spotfitter}, an ILOT based starspot modeling application, which uses a
circular, dark starspot model to find the latitude, longitude, size and
temperature of one or two spot groups on the active primary star.

We believe that knowledge of such parameter distributions will be
important for any `meta-analyses' based on collection of many individual
and separate studies deriving exoplanet parameters.  This is why we have
kept the distribution charts in this paper, and we encourage other
researchers to share their `uncertainties'. Without a clear
understanding of confidence limits in modelling results, meta-analyses
might reach incorrect conclusions by placing undue weight on
`observations' of dubious confidence or indeed the reverse.  We strongly
encourage other researchers in the field to use techniques such as MCMC
to explore the determinacy of their model fit and the confidence that
should be placed in the parameter estimates.

\section{Acknowledgements}

{ This research has made use of the NASA Exoplanet
Archive, which is operated by the California Institute of Technology,
under contract with the National Aeronautics and Space Administration
under the Exoplanet Exploration Program. } It is a pleasure to
acknowledge the additional help and encouragement from the National
University of Singapore, particularly through Prof.\ Lim Tiong Wee of
the Department of Statistics and Applied Probability. Associate
Professor Alex R.\ Cook and Associate Professor David Nott provided
valuable guidance on the Random Walk Metropolis-Hastings and Hamiltonian
Monte Carlo (HMC) applications. {\sc winfitter} is available at
http://michaelrhodesbyu.weebly.com/astronomy.html

%----------------------------------------------------------


\begin{thebibliography}{}
 
  \bibitem[\protect\citeauthoryear{akeson}{2013}]{ake} {
Akeson, R.L., Chen, X., Ciardi, D., Crane, M., Good, J., Harbut, M., 
Jackson, E., Kane, S.R., Laity, A.C., Leifer, S., Lynn, M., McElroy, D. L., 
Papin, M.,  Plavchan, P.,  Ramirez, S.V.,  Rey, R., von Braun, K., 
Wittman, M. , Abajian, M.,  Ali, B., Beichman, C.,  Beekley, A., Berriman, G.B.,
Berukoff, S.,  Bryden, G.,  Chan, B.,  Groom, S.,  Lau, C.,  Payne, A. N., 
Regelson, M.,  Saucedo, M.,  Schmitz, M.,  \& Stauffer,  J.,
Wyatt, P., \& Zhang, A., PASP, 125, 989, 2013}

 \bibitem[\protect\citeauthoryear{albert}{2014}]{albert} { 
 Albert, J., 2014,} ``LearnBayes: Functions for Learning Bayesian Inference'',
 https://CRAN.R-project.org/ package=LearnBayes
 
 
 \bibitem[\protect\citeauthoryear{Ballard}{2014}]{bal} 
 Ballard, S., et al. (25 authors), 2014, Astrophysical Journal, 790, 12
 
 \bibitem[\protect\citeauthoryear{Banks}{1990}]{ban}
 Banks, T., \& Budding, E., 1990, Ap{\&}SS, 167, 221
 
 \bibitem[\protect\citeauthoryear{Banks2}{1990}]{ban2}
  Banks, T., Sullivan, D.J., \& Budding, E., 1990, Ap{\&}SS, 173, 77
  
  \bibitem[\protect\citeauthoryear{Banks2}{1991}]{ban3}
  Banks, T., Kilmartin, P.M., \& Budding, E., 1991, Ap{\&}SS, 183, 309
  
 \bibitem[\protect\citeauthoryear{Banks3}{1994}]{ban3}
 Banks, T., Dodd, R.J., \& Sullivan, D.J., 1994, MNRAS, 272, 821  
 
  \bibitem[\protect\citeauthoryear{Banks4}{1995}]{ban4}
 Banks, T., Sullivan, D.J., \& Dodd, R.J., 1995, MNRAS, 274, 1225 
 
\bibitem[\protect\citeauthoryear{Batalha}{2014}]{batalha}
{
Batalha, N. M., 2014, PNAS, 111 (35), 12647
}
 
\bibitem[\protect\citeauthoryear{Bertsimas}{1993}]{ber} Bertsimas, D.,
\& Tsitsiklis, J., 1993,  Statistical Science, 8, 10-15

\bibitem[\protect\citeauthoryear{Bevington}{1969}]{bev} Bevington, P.R.,
1969, {\em Data Reduction and Error Analysis for the Physical Sciences},
McGraw-Hill, New York

\bibitem[\protect\citeauthoryear{Borucki}{2003}]{bor} Borucki, W.J., et
al.\ (14 authors), 2003, in {\em Scientific Frontiers in Research on
Extrasolar Planets}, Eds.\ D.\ Deming and S.\ Seager, ASP Conf.\ Ser.\,
294, 427

\bibitem[\protect\citeauthoryear{}{2011}]{bo2} Borucki, W.J., et
al.\ (69 authors), 2011, ApJ, 736, 19

 \bibitem[\protect\citeauthoryear{Brooks}{1997}]{Bro2} 
{
 Brooks, S., \& Gelman, A., 1997,
 J. Comput. Graph. Statist, 434
 }

\bibitem[\protect\citeauthoryear{Brooks}{2011}]{bro1}
Brooks, S., Gelman, A., Jones, G., \& Meng, X., 2011, {\em Handbook of Markov Chain Monte Carlo},
Chapman and Hall, CRC

\bibitem[\protect\citeauthoryear{brown}{2011}]{brown}
{
Brown, T. M.,  Latham, D. W.,  Everett, M. E., \& Esquerdo, G. A.,
2011, Ap. J, 142, 112
}

\bibitem[\protect\citeauthoryear{Budding}{2007}]{bu2} Budding, E., \&
Demircan, O., 2007, {\em Introduction to Astronomical Photometry}, CUP

\bibitem[\protect\citeauthoryear{Budding}{2016a}]{bu3} Budding, E.,
P\"{u}sk\"{u}ll\"{u}, \c{C}., Rhodes, M.D., Demircan, O., \& Erdem, A.,
2016, Ap{\&}SS, 361, 17

\bibitem[\protect\citeauthoryear{Budding}{2016b}]{bu4} Budding, E.,
Rhodes, M.D., P\"{u}sk\"{u}ll\"{u}, \c{C}., Ji, Y., Erdem, A., \& Banks,
T.,  2016, Ap{\&}SS, 361, 346

\bibitem[\protect\citeauthoryear{Charbonneau}{1999}]{ch2} Charbonneau,
D., Noyes, R.W., Korzennik, S.G., Nisenson, P., Jha, S., Vogt, S.S., \&
Kibrick, R.I., 1999, ApJ, 527, 445

\bibitem[\protect\citeauthoryear{Ciupke}{2016}]{ciupke}
{
Ciupke, K., 2016,} ``psoptim: Particle Swarm Optimization'', 
https://CRAN.R-project.org/package=psoptim


\bibitem[\protect\citeauthoryear{Cortez}{2014}]{cor} Cortez, P., 2014,
{\em Modern Optimization with R (1st Edition ed.)},  Springer,
Switzerland.


\bibitem[\protect\citeauthoryear{Csizmadia}{2013}]{ecsizmadia}
{
Csizmadia, Sz., Pasternacki, Th., Dreyer, C., Cabrera, J., Erikson, A., \& Rauer, H.,
2013, A\&A, 549, A9
}

\bibitem[\protect\citeauthoryear{Elzhov}{2016}]{elzhov}
{
Elzhov, T.V., Mullen,  K. M., Spiess, A. N, \& Bolker, B., 2016,
 ``minpack.lm: R Interface to the Levenberg-Marquardt Nonlinear
Least-Squares Algorithm Found in MINPACK, Plus Support for Bounds'',
https://CRAN.R-project.org/package=minpack.lm
}

\bibitem[\protect\citeauthoryear{Espinoza}{2015}]{esp} Espinoza, N., \&
Jordan, A., 2015, MNRAS, 450, 1879

\bibitem[\protect\citeauthoryear{Espinoza}{2016}]{esp} Espinoza, N., \&
Jordan, A., 2016, MNRAS, 457, 3573

\bibitem[\protect\citeauthoryear{Ford}{2005}]{for} Ford, E. B., 2005,
AJ, 129, 1706

\bibitem[\protect\citeauthoryear{Gandolfi}{2013}]{gandolfi}{
Gandolfi, D., Parviainen, H., Fridlund, M., Hatzes, A. P., Deeg, H. J.,
Frasca, A., Lanza, A.F., Prada Moroni, P. G., Tognelli, E., McQuillan, A.,
Aigran, S., Alonso, R., Antoci, V., Cabrera, J., Carone, L., Csizmadia, Sz., 
Djupvik, A. A., Guenther, E.W., Jessen-Hansen, J., Ofir, A., \& Telting, J.,
2013, A\&A, 557, A74
}

\bibitem[\protect\citeauthoryear{Gelman}{1992}]{gelman1}
{
Gelman, A., \& Rubin, D. B.,  1992, Statistical Science 7, 457}

\bibitem[\protect\citeauthoryear{Givens}{2011}]{givens}
{
Givens, G.H., \& Hoeting, J. A., 2013, {\em Computational Statistics - 2nd Edition'}, Wiley
}

\bibitem[\protect\citeauthoryear{Mandel}{2002}]{man} Mandel, K., Agol,
E., 2002, ApJ, 580, 171

\bibitem[\protect\citeauthoryear{Mitchell}{1998}]{mitchell} Mitchell, M.,
1998, ``An Introduction to Genetic Algorithms (Complex Adaptive
Systems)'', MIT Press

\bibitem[\protect\citeauthoryear{Moutou}{2013}]{moutou} {
Moutou, C., Deleuil, M., Guillot, T.,  Baglin, A., Bord�, P., Bouchy, F.,
Cabrera, J., Csizmadia, S., \& Deeg, H.J., 2013, Icarus, 226, 1625
}

\bibitem[\protect\citeauthoryear{Mullaly}{2015}]{mul} Mullally, F.,
Coughlin, J. L., Thompson, S. E., et al. 2015, ApJS, 217, 31

\bibitem[\protect\citeauthoryear{Nutzman}{2009}]{nutz}{
Nutzman, P., Charbonneau, D.,  Winn, J. N., Knutson, H.A., 
Fortney, J.J.,  Holman, M. J., \& Agol, E., 
Ap. J., 692, 229, 2009
}

\bibitem[\protect\citeauthoryear{Pollacco}{2006}]{pol} Pollacco, D.L.,
et al.\ (27 authors), 2006, PASP, 118, 1407

\bibitem[\protect\citeauthoryear{Pont}{2006}]{pont} 
{
Pont, F.,  Zucker, S.,  \& Queloz, D., 
{\em MNRAS},  2006,  373, 231}

\bibitem[\protect\citeauthoryear{RCore}{2014}]{rco} R Core Team, 2014,
{\em R: A language and environment for statistical computing}, R
Foundation for Statistical Computing, Vienna, Austria.

\bibitem[\protect\citeauthoryear{rhodes}{2014}]{nutz}{
Rhodes, M.D., \& Budding, E., 2014, ApSS, 351: 451
}

\bibitem[\protect\citeauthoryear{Rice}{2014}]{rice} Rice, K., 2014,
Challenges, 5, 296

\bibitem[\protect\citeauthoryear{Ricker}{2010}]{ricker}{
Ricker, G. R., Latham, D. W., Vanderspek, R. K., Ennico, K. A., Bakos, G.,  
Brown, T. M., Burgasser, A. J., Charbonneau, D., Clampin, M.,  Deming, L. D., Doty, J. P.,  
Dunham, E. W.,  Elliot, J. L., Holman, M. J., Ida, S.,  Jenkins, J. M.,  Jernigan, J. G., 
Kawai, N., Laughlin, G. P., Lissauer, J. J., Martel, F.,  Sasselov, D. D., Schingler, R. H., 
Seager, S., Torres, G., Udry, S., Villasenor, J. N., Winn, J. N., \& Worden, S. P.,
American Astronomical Society, AAS Meeting \#215, id.450.06; Bulletin of the American Astronomical Society, Vol. 42, 459}

\bibitem[\protect\citeauthoryear{Rini}{2011}]{rini} Rini, D.P.,
Shamsuddin, S.M., \& Yuhaniz, S.S., 2011, International Journal of
Computer Applications, 14, 19

\bibitem[\protect\citeauthoryear{Rowe}{2014}]{rowe} Rowe, J. F., Bryson,
S. T., Marcy, G. W., et al. 2014, ApJ, 784, 45

\bibitem[\protect\citeauthoryear{Sinharay}{2003}]{Sinh1} 
{ Sinharay, S., 2003,
{\em Assessing Convergence of the Markov Chain Monte Carlo Algorithm: A Review},
ETS Research Report Series, i-52.
}

\bibitem[\protect\citeauthoryear{Sobolev}{1975}]{sob} Sobolev, V.V.,
1975, {\em Light Scattering in Planetary Atmospheres}, Pergamon Press,
Oxford

\bibitem[\protect\citeauthoryear{Stan}{2016}]{stan}
{ 
Stan Development Team, 2016, ``RStan: the R interface to Stan'',
http://mc-stan.org/}

{
 \bibitem[\protect\citeauthoryear{Tenenbaum}{2010}]{Tenen1}
 Tenenbaum, P., Brysona, S. T., Chandrasekarana, H., Jie, L, Quintanaa, E., Twickena, J. D., \& Jenkinsa, J. M., 
 2010, {\em Proc. SPIE}, 7740
 }

\bibitem[\protect\citeauthoryear{Wang}{2013}]{wang2013} Wang, Z.,
Mohamed, S.,  \& de Freitas, N., 2013, {\em Adaptive Hamiltonian and
Riemann Manifold Monte Carlo Samplers},   In: International Conference on Machine Learning (ICML), 
JMLR W\&CP, 28(3), 1462

\bibitem[\protect\citeauthoryear{Xiang}{2013}]{xiang}
{
Xiang, Y., Gubian, S., Suomela, B., \& Hoeng, J.,  2013,
``Generalized Simulated Annealing for Efficient Global Optimization: 
the GenSA Package for R", The R
  Journal, Volume 5/1, URL http://journal.r-project.org/}

\bibitem[\protect\citeauthoryear{Zielik}{1988}]{zeilik1988} Zeilik, M.,
DeBlasi, C., Rhodes, M., \& Budding, E., 1988, Astrophys. J. 332, 293

\bibitem[\protect\citeauthoryear{Zeilik}{1994}]{zeilik1994} Zeilik, M.,
Gordon, S., Jaderlund, E., Ledlow, M. J., Summers, D. L., Heckert, P.
A., Budding, E., \& Banks, T.,  1994, Astrophys. J, 421, 303

\end{thebibliography}
 \end{document}